\documentclass[preprint, superscriptaddress, preprintnumbers, amsmath, amssymb]{revtex4}

\allowdisplaybreaks
\allowdisplaybreaks[4]

\usepackage{CJK}
\usepackage{graphicx}% Include figure files
\usepackage{dcolumn}% Align table columns on decimal point
\usepackage{bm}% bold math
\usepackage[dvipdfm,
            pdfstartview=FitH,
            CJKbookmarks=true,
            bookmarksnumbered=true,
            bookmarksopen=true,
            colorlinks=true,
            pdfborder=001,
            linkcolor=blue,
            anchorcolor=blue,
            citecolor=blue,
            ]{hyperref}

\begin{document}

\begin{CJK}{GBK}{song}

\title{Multiple chiral doublets in four-$j$ shells particle rotor model:
five possible chiral doublets in $^{136}_{~60}$Nd$_{76}$}

\author{Q. B. Chen}
\affiliation{Physik-Department, Technische Universit\"{a}t
M\"{u}nchen, D-85747 Garching, Germany}

\author{B. F. Lv}
 \affiliation{Centre de Sciences Nucl\'eaires et Sciences de la
  Mati\`ere, CNRS/IN2P3, Universit\'{e} Paris-Saclay, B\^at.~104-108, 91405 Orsay, France}

\author{C. M. Petrache}
\affiliation{Centre de Sciences Nucl\'eaires et Sciences de la
  Mati\`ere, CNRS/IN2P3, Universit\'{e} Paris-Saclay, B\^at.~104-108, 91405 Orsay, France}

\author{J. Meng}\email{mengj@pku.edu.cn}
\affiliation{State Key Laboratory of Nuclear Physics and Technology,
             School of Physics, Peking University, Beijing 100871, China}%
\affiliation{Yukawa Institute for Theoretical Physics, Kyoto
             University, Kyoto 606-8502, Japan}
\affiliation{Department of Physics, University of Stellenbosch,
             Stellenbosch, South Africa}%

\date{\today}

\begin{abstract}

A particle rotor model, which couples nucleons in four
single-$j$ shells to a triaxial rotor core, is developed
to investigate the five pairs of nearly degenerate doublet bands
recently reported in the even-even nucleus $^{136}$Nd.
The experimental energy spectra and available $B(M1)/B(E2)$
values are successfully reproduced. The angular momentum geometries of the
valence nucleons and the core support the chiral rotation interpretations
not only for the previously reported chiral doublet, but also for the other
four candidates. Hence, $^{136}$Nd is the first even-even candidate nucleus
in which the multiple chiral doublets exist. Five pairs of chiral doublet
bands in a single nucleus is also a new record in the study of nuclear
chirality.

\end{abstract}

\maketitle

%%%%%%%%%%%%%%%%%%%%%%%%%%%%%%%%%%%%%%%%%%%%%%%%%%%%%%%%%%
%                    begin  introduction
%%%%%%%%%%%%%%%%%%%%%%%%%%%%%%%%%%%%%%%%%%%%%%%%%%%%%%%%%%

Chiral rotation is an exotic rotational mode in a nucleus with triaxial
ellipsoidal shape. The rotations about an axis out of the three principal
planes of the triaxial nucleus can give rise to a pair of near degenerate $\Delta I = 1$
bands with the same parity, i.e., chiral doublet bands~\cite{Frauendorf1997NPA}.
Chiral rotation was well established in the $A \sim 80$, 100, 130, and 190 mass
regions in odd-odd nuclei~\cite{S.Y.Wang2011PLB, Vaman2004PRL, Tonev2014PRL,
Starosta2001PRL, Grodner2006PRL, Balabanski2004PRC} and odd-$A$
nuclei~\cite{Zhu2003PRL, Mukhopadhyay2007PRL, Petrache2016PRC}.
For details, see recent reviews~\cite{J.Meng2010JPG, J.Meng2014IJMPE,
Bark2014IJMPE, J.Meng2016PS, Raduta2016PPNP, Starosta2017PS, Frauendorf2018PS}
or data tables~\cite{B.W.Xiong2018arXiv}.

However, chiral doublet bands were rarely observed in even-even
nuclei. The general opinion for this is that the multi-quasiparticle
configurations become more complex and involve at least two valence
protons and two valence neutrons. In Ref.~\cite{Y.X.Luo2009PLB}, two
doublet bands were observed in the even-even isotopes $^{110,112}$Ru and
interpreted as soft chiral vibrations.

Very recently, five pairs of nearly degenerate doublet bands were reported
in even-even nucleus $^{136}$Nd, which were discovered in a high-statistics
experiment performed with the high-efficiency Jurogam II
array~\cite{Petrache2018PRC}. It was demonstrated that the
chiral partners of strongly populated bands in the triaxial nucleus are
present close to yrast, as in the case of the odd-odd and odd-even
nuclei, but are far weaker than the yrast partners and therefore not
easy to observe. The observed five pairs of nearly degenerate bands were investigated
by the constrained and tilted axis cranking covariant density functional theory
(TAC-CDFT)~\cite{J.Meng2006PRC, P.W.Zhao2011PLB, J.Meng2013FP, P.W.Zhao2017PLB, J.Meng2016book}:
one of them is revealed to be a chiral doublet, and the other four are chiral
candidates~\cite{Petrache2018PRC}. These observations shed new lights on the investigations of chiral
doublets in even-even nuclei. If the four chiral candidates are finally confirmed,
then they will constitute a multiple chiral doublet (M$\chi$D), a phenomenon predicted by
covariant density functional theory (CDFT)~\cite{J.Meng2006PRC,
J.Peng2008PRC, J.M.Yao2009PRC, J.Li2011PRC, J.Li2018PRC} and particle
rotor model (PRM)~\cite{Droste2009EPJA, Q.B.Chen2010PRC,
Hamamoto2013PRC, H.Zhang2016CPC}, and observed
experimentally~\cite{Ayangeakaa2013PRL, Kuti2014PRL, C.Liu2016PRL}.
The future identification of such bands in $^{136}$Nd will
hopefully open a campaign of measurements for other even-even
triaxial nuclei, in which the chirality or the M$\chi$D phenomenon could
exist.

Theoretically, various approaches have been developed extensively
to investigate the chiral doublet bands.
For example, the PRM~\cite{Frauendorf1997NPA, J.Peng2003PRC,
Koike2004PRL, S.Q.Zhang2007PRC, B.Qi2009PLB}, the TAC
approach~\cite{Dimitrov2000PRL, Olbratowski2004PRL, Olbratowski2006PRC,
P.W.Zhao2017PLB}, the TAC plus random-phase approximation (RPA)~\cite{Almehed2011PRC},
the collective Hamiltonian method~\cite{Q.B.Chen2013PRC, Q.B.Chen2016PRC}, the
interacting boson model~\cite{Brant2008PRC, Brant2009PRC}, and the angular momentum
projection (AMP)~\cite{Bhat2012PLB, Bhat2014NPA, F.Q.Chen2017PRC, Shimada2018PRC_v1}.

In Ref.~\cite{Petrache2018PRC}, as mentioned above, the observed doublet bands
in $^{136}$Nd were investigated in the framework of the TAC-CDFT~\cite{P.W.Zhao2011PLB, J.Meng2013FP,
P.W.Zhao2017PLB, J.Meng2016book}, which is a fully microscopic approach, but
cannot describe the energy splitting and the quantum tunneling between
the chiral doublet bands. The aim of the present work is to investigate
the chirality of $^{136}$Nd in the framework of PRM.
PRM is a quantal model coupling the collective rotation and the single-particle
motions. In contrast to the TAC approach, it describes a system in the laboratory
frame. The total Hamiltonian are diagonalized with total
angular momentum as a good quantum number, and the energy splitting and quantum
tunneling between the doublet bands can be obtained directly. Moreover,
the basic microscopic inputs for PRM can be obtained from the constrained
CDFT~\cite{J.Meng2006PRC, J.Meng2016book, Ayangeakaa2013PRL, Lieder2014PRL,
Kuti2014PRL, C.Liu2016PRL, Petrache2016PRC}.

Various versions of PRM have been developed to investigate the
chiral doublet bands with different kinds of configurations. For example,
the PRM with one-particle-one-hole configuration was used to describe
the chirality in odd-odd nuclei~\cite{Frauendorf1997NPA, J.Peng2003PRC,
Koike2004PRL, B.Qi2009PRC, H.Zhang2016CPC}. To simulate the effects of many
valence nucleons, pairing correlations were introduced and
PRM with two quasiparticles configuration was
developed~\cite{Koike2003PRC, S.Q.Zhang2007PRC, S.Y.Wang2007PRC, S.Y.Wang2008PRC,
Lawrie2008PRC, Lawrie2010PLB, Shirinda2012EPJA}. To describe the
odd-$A$ nuclei, the many-particle-many-hole versions of PRM with nucleons in two
single-$j$ shells~\cite{B.Qi2009PLB, B.Qi2011PRC} or three
single-$j$ shells~\cite{Ayangeakaa2013PRL, B.Qi2013PRC, Lieder2014PRL,
Kuti2014PRL}, have been developed. It is noted that the unpaired
nucleon configurations of the doublet bands in the even-even nucleus
$^{136}$Nd involve four different single-$j$ shells. Such
PRM is still unavailable.

In this letter, a PRM that couples nucleons in four
single-$j$ shells to a triaxial rotor core is developed and applied
to study the energy spectra, the electromagnetic transition probabilities,
as well as the angular momentum geometries for the observed
doublet bands in $^{136}$Nd.

The formalism of the PRM in the present work is an extension of
that in Ref.~\cite{B.Qi2009PLB}, where the many-particle-many-hole
version of PRM with two single-$j$ shells was developed.
The total Hamiltonian of PRM is expressed as
\begin{align}\label{eq1}
\hat{H}_\textrm{PRM}=\hat{H}_{\rm coll}+\hat{H}_{\rm intr},
\end{align}
with the collective rotor Hamiltonian
\begin{align}
\hat{H}_{\rm coll}=\sum_{k=1}^3 \frac{\hat{R}_k^2}{2\mathcal{J}_k}
 =\sum_{k=1}^3 \frac{(\hat{I}_k-\hat{J}_k)^2}{2\mathcal{J}_k},
\end{align}
where the indexes $k=1$, 2, and 3 refer to the three principal axes of
the body-fixed frame. The $\hat{R}_k$ and $\hat{I}_k$ denote the
angular momentum operators of the core and of the total nucleus, respectively,
and the $\hat{J}_k$ is the total angular momentum operator of the
valence nucleons. The moments of inertia of the irrotational
flow type~\cite{Ring1980book} are adopted, i.e.,
$\mathcal{J}_k=\mathcal{J}_0\sin^2(\gamma-2k\pi/3)$, with
$\gamma$ the triaxial deformation parameter. In addition,
the intrinsic Hamiltonian is written as
\begin{align}
 H_{\rm intr}=\sum_{i=1}^4 \sum_{\nu} \varepsilon_{i, \nu} a_{i,\nu}^\dag a_{i,\nu},
\end{align}
where $\varepsilon_{i, \nu}$ is the single particle energy
in the $i$-th single-$j$ shell provided by
\begin{align}\label{eq2}
 h_{\rm sp}=\pm \frac{1}{2}C\Big\{\cos \gamma\big(j_3^2-\frac{j(j+1)}{3}\big)
                +\frac{\sin \gamma}{2\sqrt{3}}\big(j_+^2+j_-^2\big)\Big\}.
\end{align}
Here, the plus or minus sign refers to particle or hole, and the
coefficient $C$ is proportional to the quadrupole deformation $\beta$
as in Ref.~\cite{S.Y.Wang2009CPL}.

The single particle state and its time reversal state are expressed as
\begin{align}
 a_\nu^\dag|0\rangle & =\sum_{\alpha\Omega} c_{\alpha\Omega}^{\nu}|\alpha,j\Omega\rangle,\\
 a_{\bar{\nu}}^\dag|0\rangle &=\sum_{\alpha\Omega} (-1)^{j-\Omega} c_{\alpha\Omega}^{\nu}|\alpha,j-\Omega\rangle,
\end{align}
where $\Omega$ is the projection of the single-particle angular momentum
$j$ along the 3-axis of the intrinsic frame and
restricted to $\dots$, $-3/2$, 1/2, 5/2, $\dots$ due to
the time-reversal degeneracy, and $\alpha$ denotes the other
quantum numbers. For a system with $\sum_{i=1}^4 N_i$ valence
nucleons ($N_i$ denotes the number of the nucleons in the $i$-th
single-$j$ shell), the intrinsic wave function is given as
\begin{align}
 |\varphi\rangle=\prod_{i=1}^4 \Big(\prod_{l=1}^{n_{i}} a_{i,\nu_l}^\dag \Big)
  \Big(\prod_{l=1}^{n_{i}^\prime} a_{i,\bar{\mu}_l}^\dag \Big)|0\rangle,
\end{align}
with $n_{i}+n_{i}^\prime=N_i$ and $0\leq n_{i} \leq N_i$.

The total wave function can be expanded into the strong coupling
basis
\begin{align}\label{eq3}
 |IM\rangle=\sum_{K\varphi} c_{K\varphi}|IMK\varphi\rangle,
\end{align}
with
\begin{align}
 |IMK\varphi\rangle
  =\frac{1}{\sqrt{2(1+\delta_{K0}\delta_{\varphi,\bar{\varphi}})}}
   \big(|IMK\rangle|\varphi\rangle
  +(-1)^{I-K}|IM-K\rangle|\bar{\varphi}\rangle\big),
\end{align}
where $|IMK\rangle$ is the Wigner function $\sqrt{\frac{2I+1}{8\pi^2}}D_{MK}^I$.
The basis states are symmetrized under the point group $\textrm{D}_2$, which
leads to $K-\frac{1}{2}\sum_{i=1}^4 (n_{i}-n_{i}^\prime)$ being an even
integer.

It is noted that due to the inclusion of many-particle-many-hole configurations
with four single-$j$ shells, the size of the basis space is rather large. For
example, for the calculations of band D1 in $^{136}$Nd (see its configuration in
Table~\ref{tab1}), the dimension of the basis space is $864(2I+1)$. For $I=10\hbar$,
this value is 18144, and for $I=20\hbar$, it is 35424. It is quite time-consuming
in the diagonalization of the PRM Hamiltonian matrix. To solve this
problem, similar in the shell-model-like approach (SLAP)~\cite{J.Y.Zeng1983NPA, Z.Shi2018PRC},
we adopt a properly truncated basis space by introducing a cutoff for the
configuration energy $\sum_{i,\nu} \varepsilon_{i,\nu}$. In such a way, the
dimension of the PRM matrix is reduced to $\sim 5000$-$10000$ with the energy
uncertainty within 0.1$\%$.

After obtaining the wave functions of PRM, the reduced transition probabilities
$B(M1)$ and $B(E2)$, and expectation values of the angular momentum of the system
can be calculated.

There are five pairs of doublet bands in $^{136}$Nd (labeled as bands
D1-D5), in which three of them (bands D1, D2, and D5) are positive parity.
Besides, there is a dipole band (labeled as band D6), which has no partner band.
In the PRM calculations for these bands, the unpaired nucleon configurations are
consistent with those in Ref.~\cite{Petrache2018PRC} and the corresponding quadrupole
deformation parameters $(\beta,\gamma)$ are obtained from triaxial constrained CDFT
calculations~\cite{J.Meng2006PRC}. The moments of inertia $\mathcal{J}_0$ and
Coriolis attenuation factors $\xi$ are adjusted to reproduce the trend of
the energy spectra. The corresponding details are listed in Table~\ref{tab1}. In
addition, for the electromagnetic transitions, the empirical
intrinsic quadrupole moment $Q_0=(3/\sqrt{5\pi})R_0^2Z\beta$, and
gyromagnetic ratios for rotor $g_R=Z/A$ and for nucleons $g_{p(n)}=g_l+(g_s-g_l)/(2l+1)$
($g_l=1(0)$ for protons (neutrons) and $g_s=0.6g_{s}(\textrm{free})$)~\cite{Ring1980book}
are adopted.

\begin{table*}[!ht]
  \centering
  \caption{The parities, unpaired nucleon configurations, quadrupole deformation parameters
  ($\beta$, $\gamma$), moments of inertia $\mathcal{J}_0$ (unit $\hbar^2/\textrm{MeV}$),
  and Coriolis attenuation factors $\xi$ used in the PRM calculations for bands D1-D6 and
  their partners.}
  \label{tab1}
  \begin{tabular}{cccccc}
\hline\hline
Band & Parity & Unpaired nucleons & ($\beta$,$\gamma$) & $\mathcal{J}_0$ & $\xi$\\
\hline
D1& $+$ & $\pi(1h_{11/2})^1(2d_{5/2})^{-1}\otimes\nu (1h_{11/2})^{-1}(2d_{3/2})^{-1}$
&~~($0.21$, $21^\circ$)~~ & ~~32.0~~ & ~~0.96~~\\
D2& $+$ & $\pi(1h_{11/2})^3(2d_{5/2})^{-1}\otimes\nu (1h_{11/2})^{-1}(2d_{3/2})^{-1}$
&($0.22$, $19^\circ$) & 35.0 & ~~0.96~~\\
D5& $+$ & $\pi(1h_{11/2})^2(1g_{7/2})^{-2}\otimes\nu (1h_{11/2})^{-1}(1f_{7/2})^1$
&($0.26$, $23^\circ$) & 40.0 & ~~0.93~~\\
D6& $+$ & $\pi(1h_{11/2})^3(2d_{5/2})^{-1}\otimes\nu (1h_{11/2})^{-1}(1f_{7/2})^1$
&($0.23$, $25^\circ$) & 42.0 & ~~0.95~~\\
\hline
D3& $-$ & $\pi(1h_{11/2})^2\otimes\nu(h_{11/2})^{-1}(2d_{3/2})^{-1}$
& ($0.22$, $19^\circ$) & 32.0 & ~~0.97~~\\
D4& $-$ & $\pi(1h_{11/2})^2(2d_{5/2})^{-2}\otimes\nu(h_{11/2})^{-1}(2d_{3/2})^{-1}$
& ($0.22$, $19^\circ$) & 33.0 & ~~0.97~~\\
\hline
  \end{tabular}
\end{table*}

\begin{figure}[!th]
  \begin{center}
    \includegraphics[width=8 cm]{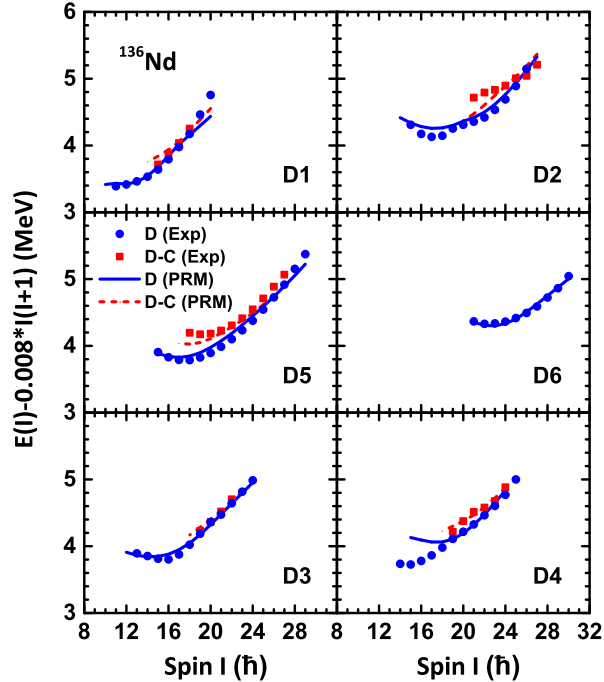}
    \caption{(Color online) The energy spectra of bands D1-D6 and their partners
    calculated by PRM in comparison with corresponding data. The excitation energies
    are relative to a rigid-rotor reference.}\label{fig1}
  \end{center}
\end{figure}

The calculated energy spectra for the bands D1-D6 in $^{136}$Nd are presented in
Fig.~\ref{fig1}, together with the corresponding data. The experimental energy
spectra are reproduced excellently by the PRM calculations. Being a quantum model,
PRM is able to reproduce the energy splitting for the whole observed
spin region. It is seen that except for D2, the trend and amplitude
for the energy splitting between partner bands are reproduced well.

For D2 and D5, the energy differences between the doublet
bands decrease gradually with the spin. In detail, for D2, the
energy splitting is $\sim 360~\textrm{keV}$ at $I=21\hbar$, and finally goes
to $\sim 110~\textrm{keV}$ at $I=25\hbar$. The PRM calculations underestimate
this energy separation. For bands D5 and D5-C, which are identified as chiral doublets in
Ref.~\cite{Petrache2018PRC}, their energy splitting is $\sim 410~\textrm{keV}$
at $I=18\hbar$, and reduces to $\sim 150~\textrm{keV}$.

For the other three pairs of double bands D1, D3, and D4, the energy separations
are about $70$, $40$, and $120~\textrm{keV}$, respectively,
and do not change much with the spin. For band D4, the PRM
calculations can not reproduce the data below $I=18\hbar$,
indicating that the used configuration is not suitable for lower spin part.
Nevertheless, the rather small energy differences
between these doublets support that they are chiral candidates.

\begin{figure}[!th]
  \begin{center}
    \includegraphics[width=8 cm]{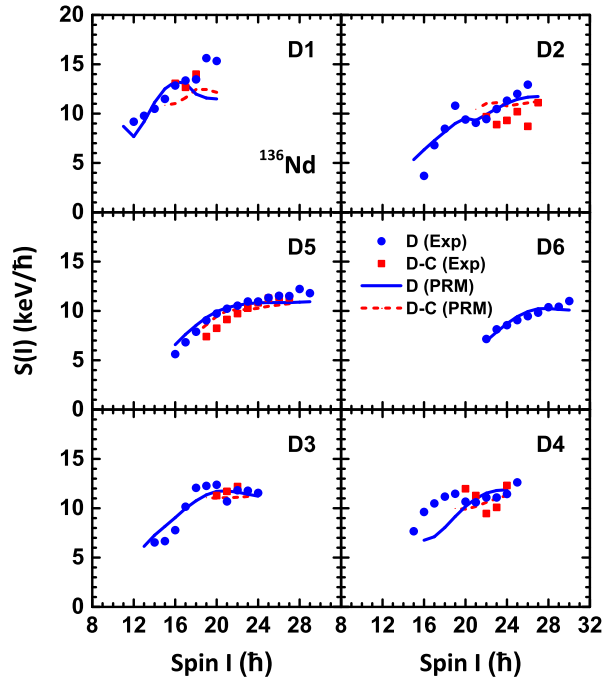}
    \caption{(Color online) The staggering parameters of bands D1-D6
    calculated by PRM in comparison with corresponding data.}\label{fig6}
  \end{center}
\end{figure}

From the energy spectra, the staggering parameters $S(I)=[E(I)-E(I-1)]/2I$
are extracted and displayed in Fig.~\ref{fig6}. The standard fingerprints for
chiral bands outlined in Ref.~\cite{Vaman2004PRL} require that $S(I)$ is
independent of spin. Overall, the PRM calculations can reproduce the behaviors
of experimental $S(I)$. Moreover, the $S(I)$ of all bands vary smoothly and do
not change much with spin. These phenomena further provide the support that the
bands D1-D5 are chiral partners.

\begin{figure}[!th]
  \begin{center}
    \includegraphics[width=8 cm]{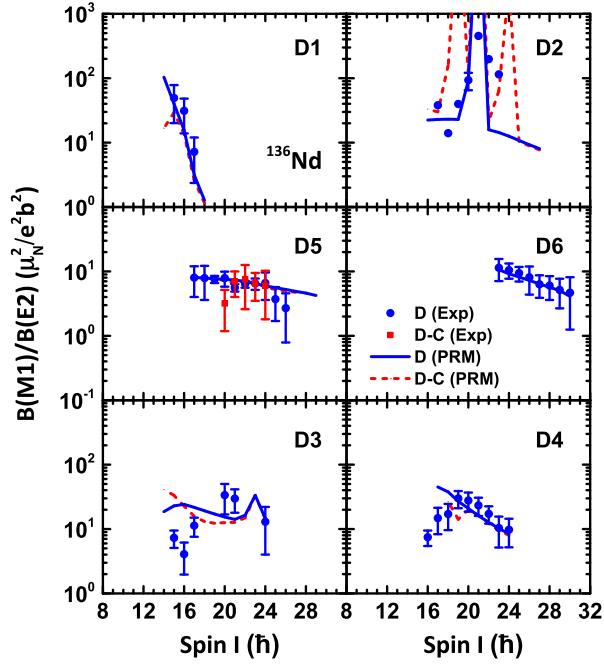}
    \caption{(Color online) The $B(M1)/B(E2)$ of bands D1-D6 and their partners
    calculated by PRM in comparison with corresponding data.}\label{fig2}
  \end{center}
\end{figure}

In Fig.~\ref{fig2}, the $B(M1)/B(E2)$ values of bands D1-D6 calculated by
PRM in comparison with corresponding data available are shown. One observes that
the PRM calculations show an impressive agreement with the data. Moreover,
excepting band D2 the calculated $B(M1)/B(E2)$ values of the
doublet bands are rather similar.

For bands D1 and D6, the $B(M1)/B(E2)$ values decrease with spin.
For D2, an abrupt increase of $B(M1)/B(E2)$ is observed at $I=21\hbar$.
The large calculated $B(M1)/B(E2)$ value at $I=21\hbar$ results from
the small $B(E2)$ value. After analyzing the corresponding PRM wave function,
we find that, at $I<20\hbar$, the largest component of the state is
$I_s \sim I$ ($I_s$ the angular momentum component
along the short axis), while for $I\geq 20\hbar$, $I_s
\sim I-2$. This structure change causes the small $B(M1)$ value at
$I=20\hbar$ and small $B(E2)$ values at $I=20$ and $21\hbar$, and hence
large $B(M1)/B(E2)$ value at $I=21\hbar$. For D5 and D5-C,
their $B(M1)/B(E2)$ values are quite similar and fulfil the
characteristics of chiral doublet bands~\cite{Koike2004PRL, B.Qi2009PRC}.
Therefore, they were identified as chiral doublet bands in Ref.~\cite{Petrache2018PRC}.

For bands D3 and D4, their $B(M1)/B(E2)$ values are
similar. They exhibit a trend that first increase and then decrease
with increasing spin. Moreover, their quasi-particle alignments show
pronounced similarity in a wide interval of rotational frequency,
shown in Ref.~\cite{Petrache2018PRC}. It seems that they were
a M$\chi$D built on identical configuration as in
$^{103}$Rh~\cite{Kuti2014PRL}. However, as their spectra
are interweaved each other at several spins, this possibility is excluded.
In the calculations, we use a configuration with three
single-$j$ shells to describe D3 and a configuration
with four single-$j$ shells to describe D4, shown in Table~\ref{tab1}.
Admittedly, the present PRM calculations do not agree very well with
the data of D3. For D4, the calculated results reproduce very well
the experimental data for $I \geq 19\hbar$.

Note that only the $B(M1)/B(E2)$ ratios for the doublet
bands D5 and D5-C has been measured. Hence, the other four pairs of
doublet bands are considered only as chiral candidates. As mentioned
before, the calculated $B(M1)/B(E2)$ values are similar in the doublet
bands. This suggests that the other four candidates might also
be chiral doublets. Definitely, further experimental efforts are
highly demanded to obtain solid evidence for the chiral doublets
interpretation.

\begin{figure}[!th]
  \begin{center}
    \includegraphics[width=10.5 cm]{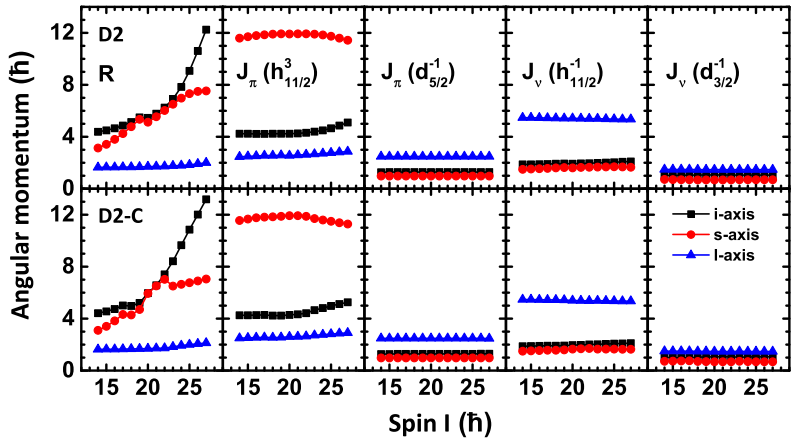}
    \caption{(Color online) The root mean square components along the intermediate
    ($i$-, squares), short ($s$-, circles) and long ($l$-, triangles) axes of the
    rotor, valence protons, and valence neutrons angular momenta calculated as
    functions of spin by PRM for the doublet bands D2 and D2-C in $^{136}$Nd.}\label{fig3}
  \end{center}
\end{figure}

\begin{figure}[!ht]
  \begin{center}
    \includegraphics[width=10.5 cm]{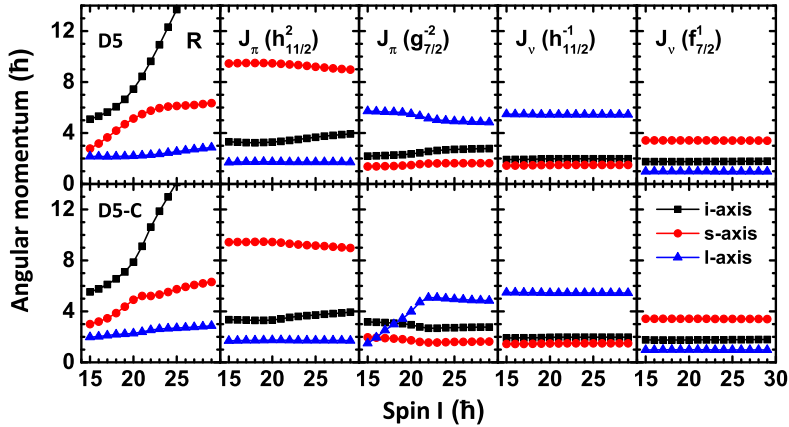}
    \caption{(Color online) Same as Fig.~\ref{fig3}, but for D5.}\label{fig5}
  \end{center}
\end{figure}

\begin{figure}[!ht]
  \begin{center}
    \includegraphics[width=10.5 cm]{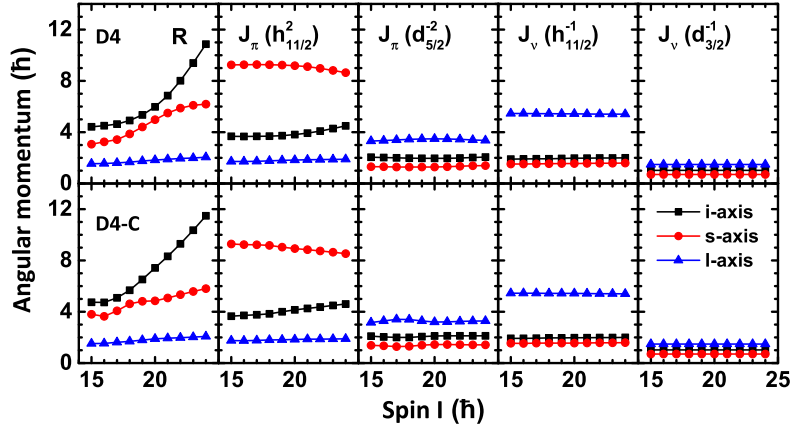}
    \caption{(Color online) Same as Fig.~\ref{fig3}, but for D4.}\label{fig4}
  \end{center}
\end{figure}

The successes in reproducing the energy spectra and electromagnetic
transition probabilities for the doublet bands in $^{136}$Nd motivate us
to examine the angular momentum geometries
of the observed bands. For this purpose, we calculate the expectation
values of the squared angular momentum components along the intermediate
($i$-), short ($s$-), and long ($l$-) axes for the rotor, valence
protons, and valence neutrons. Here, the obtained results of bands D2, D5,
and D4 are shown in Figs.~\ref{fig3}, \ref{fig5}, and \ref{fig4}
as examples, respectively.

As shown in Fig.~\ref{fig3}, for both bands D2 and D2-C, the
collective core angular momentum mainly aligns along the $i$-axis
at $I\geq 25\hbar$, because it has the
largest moment of inertia. It should be mentioned
that the $s$-component of the collective core angular
momentum is large and cannot be neglected. Moreover,
it exhibits a discontinuous behavior between $I=19$ and $20\hbar$
in the D2, and between $I=17$ and $18\hbar$ and $I=22$ and $23\hbar$ in D2-C.
This is understood as the reason of abrupt increases of $B(M1)/B(E2)$ values,
as discussed previously. The angular momentum of the three $h_{11/2}$
valence proton particles mainly aligns along the $s$-axis, and those of
valence proton and neutron holes mainly along the $l$-axis. Such orientations
form the chiral geometry of aplanar rotation. But it should be noted
that due to the large $s$-component of the rotor and proton, the total
angular momentum lies close to the $s$-$i$ plane.

For the chiral doublet bands D5 and D5-C, as shown in Fig.~\ref{fig5}, the
angular momenta have similar orientation at $I\geq 21\hbar$,
as those in D2. Namely, the angular momentum of the rotor mainly
aligns along the $i$-axis, the two $h_{11/2}$ valence proton and one $f_{7/2}$
valence neutron particles mainly align along the $s$-axis, and two $g_{7/2}$
valence proton and one neutron $h_{11/2}$ valence holes mainly along
the $l$-axis. In comparison with those in D2, the $s$-axis components
of the angular momenta of the rotor and $h_{11/2}$ valence proton
particles in D5 are about $2\hbar$ smaller. Such orientations
form a better chiral geometry of aplanar rotation than
that of D2. At $I\leq 21\hbar$, the $l$-axis
component of angular momenta of two $g_{7/2}$ valence proton holes are
different in bands D5 and D5-C. For D5, the two proton holes are aligned
and contribute $\sim 5\hbar$. However, for D5-C, the alignment
happens when the spin increases from $17\hbar$ to $21\hbar$. At $I=17\hbar$, the two
proton holes contribute angular momenta $\sim 2\hbar$.
At $I=21\hbar$, the two proton holes contribute $\sim 5\hbar$.
Such difference causes the energy difference
between the doublet bands at this spin region
$\sim 400~\textrm{keV}$ as shown in Fig.~\ref{fig1}.

For the bands D4 and D4-C, as shown in Fig.~\ref{fig4}, similar aplanar
orientation of the angular momenta of the rotor, the particles,
and the holes can be observed. This supports that D4 and D4-C are
chiral doublets. As discussed previously, there is a band-crossing at $I=19\hbar$,
and the adopted configuration is only suitable for describing the data above
band-crossing. One observes that at $I\geq 19\hbar$, the angular momenta of
the two $h_{11/2}$ valence proton particles tend to align along $i$-axis.
This leads the increase of $B(E2)$, and hence the decrease of $B(M1)/B(E2)$
with the spin as shown in Fig.~\ref{fig2}.

In summary, a PRM coupling nucleons in four single-$j$ shells
to a triaxial rotor core is developed to investigate the five pairs of nearly
degenerate doublet bands recently reported in the even-even nucleus $^{136}$Nd.
The configurations and corresponding quadrupole deformation parameters
($\beta$, $\gamma$) are obtained from the constrained CDFT calculations.
The experimental energy spectra and available $B(M1)/B(E2)$
values are successfully reproduced. The angular momentum geometries of the
valence nucleons and the core support the chiral rotation interpretations
not only for the previously reported chiral doublet, but also the other
four candidates. Therefore, $^{136}$Nd is the first even-even candidate
nucleus in which the M$\chi$D exists. Five pairs of chiral doublet bands in a
single nucleus is also a new record in the study of nuclear chirality.
Further experimental efforts are highly encouraged to obtain solid evidence for
M$\chi$D interpretations.

This work was partly supported by the National Key R$\&$D Program of China
(Contract No. 2018YFA0404400), the Deutsche
Forschungsgemeinschaft (DFG) and National Natural Science Foundation
of China (NSFC) through funds provided to the Sino-German CRC 110
``Symmetries and the Emergence of Structure in QCD'', and the
NSFC under Grants No.~11335002 and No.~11621131001.

%%%%%%%%%%%%%%%%%%%%%%%%%%%%%%%%%%%%%%%%%%%%%%%%%%%%%%%%
%                  begin refereee
%%%%%%%%%%%%%%%%%%%%%%%%%%%%%%%%%%%%%%%%%%%%%%%%%%%%%%%%

\end{CJK}

\end{document}